\documentclass{webofc}
\usepackage[varg]{txfonts}   % Web of Conferences font
\usepackage{caption}  % <-- Make sure this line is in your preamble

\begin{document}
\title{Constructing the Hyper-Kamiokande Computing Model in the Build Up to Data Taking}

\author{\firstname{Sophie} \lastname{King}\inst{1}\fnsep\thanks{\email{sophie.king@kcl.ac.uk}} on behalf of the Hyper-Kamiokande Collaboration }
\institute{King's College London, UK}

%%%%%%%%%%%%%%%%%%%%%%%%%
%%%%%%%%%%%%%%%%%%%%%%%%%
\abstract{
%%%%%%%%%%%%%%%%%%%%%%%%%
%%%%%%%%%%%%%%%%%%%%%%%%%
Hyper-Kamiokande is a next-generation multi-purpose neutrino experiment with a primary focus on constraining CP-violation in the lepton sector.  It features a diverse science programme that includes neutrino oscillation studies, astrophysics, neutrino cross-section measurements, and searches for physics beyond the standard model, such as proton decay. 
Building on its predecessor, Super-Kamiokande, the Hyper-Kamiokande far detector has a total volume approximately 5 times larger and is estimated to collect nearly 2\,PB of data per year.
%Building on its predecessor, Super-Kamiokande, the Hyper-Kamiokande far detector has a total (fiducial) volume approximately 5 (8) times larger and is estimated to collect nearly 2\,PB of data per year.
The experiment will also include both on- and off-axis near detectors, including an Intermediate Water Cherenkov Detector. To manage the significant demands relating to the data from these detectors, and the associated Monte Carlo simulations for a range of physics studies, an efficient and scalable distributed computing model is essential. This model leverages Worldwide LHC Grid computing infrastructure and utilises the GridPP DIRAC instance for both workload management and for file cataloguing.  In this report we forecast the computing requirements for the Hyper-K experiment, estimated to reach around 35\,PB (per replica) and 8,700 CPU cores ($\sim$100,000 HS06) by 2036.  We outline the resources, tools, and workflow in place to satisfy this demand.

}
\maketitle
%

%%%%%%%%%%%%%%%%%%%%%%%%
%%%%%%%%%%%%%%%%%%%%%%%%
\section{Introduction}
\label{sec:intro}
%%%%%%%%%%%%%%%%%%%%%%%%
%%%%%%%%%%%%%%%%%%%%%%%%

The Hyper-Kamiokande (Hyper-K) experiment~\cite{bib:hkcdr, bib:hksnowmass2021} is the successor to the highly successful and accomplished Super-Kamiokande (SK)~\cite{bib:skexp, bib:skosc1998} and T2K (Tokai-to-Kamioka)~\cite{bib:t2kexp, bib:t2knature} experiments.  Currently under construction, the Hyper-K far detector (HKFD) is a 258\,kton underground water Cherenkov detector.  The inner detector will house 20,000 inward-facing photomultiplier tubes (PMTs), 50\,cm in diameter, along with a thousand multi-PMTs that each contain nineteen 8\,cm PMTs.  This region will be encompassed by the outer detector, which is designed to be instrumented with a few thousand 8\,cm PMTs to veto incoming backgrounds.  This amounts to tens of thousands of readout channels, and a post-trigger rate of around 5\,TB/day of data to be transferred and stored off-site, setting the scale for Hyper-K storage requirements.

Hyper-K will search for CP-violation in the lepton sector through the study of neutrino oscillations in an accelerator-based long-baseline neutrino oscillation configuration.  For this purpose, in addition to the far detector, Hyper-K features several near detectors, which will constrain systematic uncertainties relating to the flux and neutrino interaction models and measure the neutrino beam properties, as well as performing cross-section measurements and searches for new physics.  %The suite of near detectors also includes the T2K detector INGRID, which primarily measures the on-axis neutrino beam properties.
%; since the computing requirements of INGRID are negligible with respect to the overall estimations for Hyper-K, they are neglected in this report.  
Hyper-K is also building the Intermediate Water Cherenkov Detector (IWCD), providing a near detector that utilises the same technology and target as the far detector, as well as the ability to profile the beam across a continuous range of off-axis angles.  

The computing predictions that cover the raw and processed data needs of Hyper-K detectors, as well as the Monte Carlo (MC) production samples covering the signal, background and control samples needed for all physics analyses, are presented in Section~\ref{sec:hkCompForecasts}.  The infrastructure, tools and workflow that define the Hyper-K computing model, constructed to meet these needs in an organised and efficient manner, is defined in Section~\ref{sec:hkCompModel}.

%
%The Hyper-K computing model is complemented by a modular database and web application server and a software framework that serves both the near and far water Cherenkov detectors, ensuring that Hyper-K is well equipped to meet data processing and Monte Carlo production requirements for each of its detectors and covering all components of its physics programme.

%%%%%%%%%%%%%%%%%%%%%%%%
%%%%%%%%%%%%%%%%%%%%%%%%
\section{Hyper-K Computing Forecasts}
\label{sec:hkCompForecasts}
%%%%%%%%%%%%%%%%%%%%%%%%
%%%%%%%%%%%%%%%%%%%%%%%%

The Hyper-K far detector is currently under construction, and the experiment will start taking data in 2027.  The computing forecasts were split into two stages: the construction period, from 2023 up till the end of 2026\footnote{It should be noted that construction began in 2021, but these estimates cover the period starting in 2023 as this is the point where our MC production campaigns are starting to ramp up to a more significant level.}; and the first ten years of operation, from the start of 2027 to the end of 2036\footnote{It is expected that Hyper-K will continue for an additional 10 years till the end of 2046.  A \textit{very} rough prediction for the end of this period can be considered by doubling the 2036 estimate.}.  Internally within the Hyper-K Computing Group, a detailed year-by-year prediction has been forecast for each detector and for each sample type, with input from the Computing, Software, DAQ, Calibration and Physics working groups.  However, due to the large uncertainties on these estimates - stemming from both the current phase of rapid software development within Hyper-K, which will stabilize over the next couple of years, and because aspects of the DAQ are still being finalised - we present only the 2023 and 2036 points, and use a straight line to convey the order of magnitude while avoiding the impression of year-to-year precision. The detailed breakdown is available upon request to the computing and funding agencies that support Hyper-K.

\subsection{Aggregated CPU and Storage Projections}
\begin{figure}[h]
\centering
\captionsetup{justification=centering}
\includegraphics[width=10.8cm,clip]{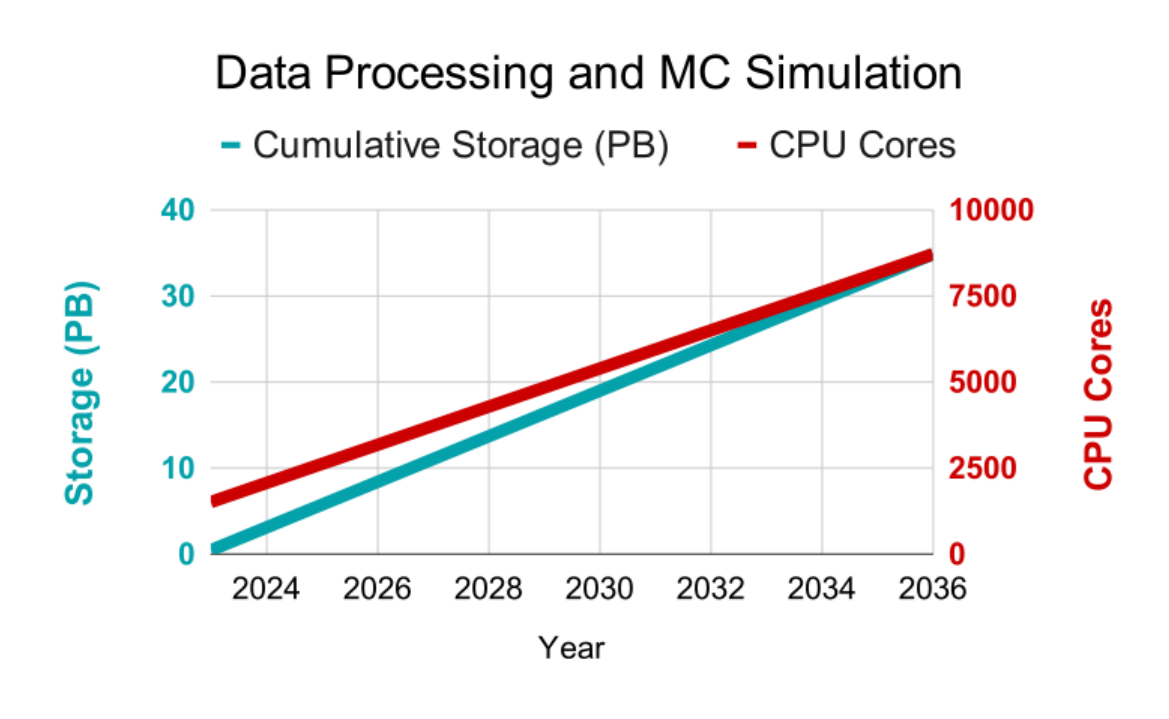}
\caption{The total computing forecasts, covering all the Hyper-K detectors, considering both data and MC needs, and for all signal, background and control/calibration samples.  Note that the storage estimate is per replica.}
\label{fig:forecast}       % Give a unique label
\end{figure}

%\textbf{Computing forecast overview:}  
The Hyper-K computing requirements for raw data storage as well as both data and MC production, across all detectors, are totalled and presented as a function of time in Figure~\ref{fig:forecast}.  The left axis indicates storage, which is projected to reach around 35\,PB for a single replica of each file.  The policy is to maintain three replicas of raw data, one in Japan and two in different countries, and two replicas of MC.  The right axis represents the number of CPU cores and is predicted to reach nearly 8,700 cores ($\sim$100,000 HS06).

%%%%%%% FIGURE %%%%%%%%%%

Throughout the construction period, the computing estimates revolve solely around the MC simulations that are generated for physics sensitivity studies, detector design optimization, validation of trigger algorithms, and calibration studies.  During this period, the IWCD MC production dominates the CPU demand and the HKFD MC dominates the storage.
%due to the production of sets of Supernova models.  

During the operational phase,  the storage and CPU associated with raw data storage, data (re)processing, calibration and event reduction/pre-selection contribute to the computing requirements.  While the HKFD raw data quickly dominates the storage needs of the experiment, accounting for around 70\% of the total, the IWCD MC production continues to dominate the CPU and is responsible for around 70\% of the total.

%\subsection{Comments on the Largest Contributors to Resource Forecasts}

\subsection{Resource-intensive tasks}

In this section we comment on the tasks that drive the CPU and storage estimates during the operational phase of the experiment.  This helps us to understand the best way to meet these demands and consider areas where we have potential to improve efficiency.

\textbf{CPU usage for Water Cherenkov reconstruction:}  The CPU requirements of the IWCD MC production, which dominates both the construction and operational periods of the experiment, is due to the reconstruction stage. While the two Water Cherenkov detectors (HKFD, IWCD) use the same reconstruction algorithm, 
%the CPU is dominated by the IWCD due to the differing detector geometry and configuration.  IWCD is much smaller than HKFD, but is instrumented with only mPMTs to retain good spatial and timing resolutions, which comes at a cost computationally.  
the close proximity of the IWCD to the neutrino beam results in a much larger number of beam events and hence requires MC samples with higher statistics.  The current reconstruction algorithm, fiTQun~/cite{bib:fitqun}, was successfully introduced for SK to reduce the fiducial volume while also improving reconstruction capabilities, and is also used within both T2K and Hyper-K.  It uses a maximum-likelihood based algorithm that computes the likelihood for different particle topologies
%the number and identification of particles 
along with their reconstructed kinematics.  Efforts are being made to improve reconstruction processing times, both by improving the efficiency of the existing method,  and by investigating  machine learning based techniques, which seek to improve not only speed but also reconstruction capabilities~\cite{bib:hkml1, bib:hkml2}.

%\textbf{Storage for HKFD MC production:}  The HKFD MC requirements dominate the storage during construction, with the supernova simulations  contributing the most.  While the supernova production is relatively fast to produce - low energy samples use a different, fast reconstruction algorithm - it requires a sizable amount of storage due to the combination of models that are simulated to satisfy DAQ stress tests as well as physics sensitivity studies.  During operation, the MC simulation for calibration studiesalso becomes significant, though it is worth noting that operational period the total HKFD MC storage is quickly eclipsed by the demands of real data.

\textbf{Raw data storage:}
The vast size of the far detector, coupled with tens of thousands of readout channels, yields a predicted (post-trigger) raw data rate of nearly 2\,PB/year to be transferred and stored off-site.
While this estimate assumes a basic level of data rejection due to triggering, it is a slightly conservative number and has the potential to decrease once the triggering is finalised and the collaboration reaches agreement about what data to store long-term.  The majority of this data will be archived to tape and brought onto disk upon request; regularly accessed data samples, based on triggers and reduction/pre-selection cuts, will be replicated to disk storage for continued ease of access.

\subsection{Future Projections}
%\textbf{Outlook}

As previously mentioned, machine learning based techniques are under investigation to improve both processing speeds and the performance of kinematic reconstruction and particle identification.  Other stages of production for which machine learning methods are being explored include simulation and calibration, albeit with lower priority.  These developments could significantly alter the Hyper-K computing needs, introducing a need for GPU-based jobs within the computing workflow and resource allocations.  These innovations could drastically reduce processing times, though it is expected that some level of likelihood-based processing would remain in conjunction with the new methods, at least for validation purposes.  While these considerations are under internal discussion, they are not included in the current computing projections.  As the timeline and performance metrics become more refined, we plan to update the computing forecasts accordingly.  

The current computing estimates focus solely on the needs concerning the storage and production of real data and MC simulations.  Internal discussions between the computing and analysis groups are underway to understand the computing requirements of the different high-level analysis frameworks.  As neutrino physics enters a systematically dominated era, statistical analysis in increasingly high-dimensional space can no longer be performed on small institute clusters.  Physics groups are instead turning to HPC clusters, with many of these analysis frameworks utilising GPU acceleration.  As the data grows in size and the complexity of these analyses increases, resource limitations can lead to the need to introduce approximations that improve speed but decrease accuracy, or to compromise on which studies can be performed.  Preparing results for significant conferences can also impose strict deadlines, resulting in large spikes in demand.  Integrating analysis level tasks into the computing framework can help to prioritise at the expense of less urgent tasks, and including the computing requirements in negotiations with computing and funding bodies can help to better plan for and guarantee these needs.  In future iterations, the computing projections and workflows for computationally demanding analysis will be presented alongside the needs of real data and MC simulation.

%%%%%%%%%%%%%%%%%%%%%%%%
%%%%%%%%%%%%%%%%%%%%%%%%
\section{The Hyper-K Computing model}
\label{sec:hkCompModel}
%%%%%%%%%%%%%%%%%%%%%%%%
%%%%%%%%%%%%%%%%%%%%%%%%

%Hyper-K adopts a tiered system for computing, similar to that of the Worldwide LHC Computing Grid (WLCG), and benefiting from much of its infrastructure and tools.  The infrastructure and tier definitions are outlined in Section~\ref{sec:hkCompTiers}. The following section, Section~\ref{sec:hkCompTools} goes on to discuss the tools we use to interact with these resources.  

Hyper-K adopts a tiered system for computing, similar to that of the Worldwide LHC Computing Grid (WLCG), and benefits from much of its infrastructure and tools.  To meet specific requirements of Hyper-K jobs and data management, we utilize community-based tools which we integrate with custom Hyper-K software.  This is discussed in Section~\ref{sec:hkCompTools}, while the tier definitions are outlined Section~\ref{sec:hkCompTiers}.

%%%%%%%%%%
\subsection{Computing Tools, Production Workflow and Data Management}
\label{sec:hkCompTools}
%%%%%%%%%%

\subsubsection{Distributed Computing with GridPP DIRAC}
\label{sec:dirac}
The DIRAC (Distributed Infrastructure with Remote Agent Control) project \cite{bib:dirac} is a software framework that interfaces users with computing resources.  It offers a pilot-based Workload Management System (WMS) for job submission, which can connect to grid, cloud and batch systems.  DIRAC also provides a Data Management System (DMS) for cataloging file replication and metadata, in the Dirac File Catalogue (DFC), along with and the associated end-user tools to manage this.  Hyper-K uses both of these services, provided by the GridPP instance of DIRAC hosted at Imperial College London~\cite{bib:gridppdirac}.  This is a multi-virtual organisation (multi-VO) service that Hyper-K accesses through the hyperk.org VO.

All T1 and T2 Hyper-K storage is managed through the DIRAC DMS and DFC.  Efficient data transfer between sites is possible through third party services such as FTS3.  For bulk data operations Hyper-K uses FTS3 with the Dirac Request Management System (RMS), which provides a scheduling and monitoring service.

At the time of writing, Hyper-K uses DIRAC for job submission to access both allocated and opportunistic resources at UK (RAL, ICL and other T2 sites), Italian (INFN) and French (IN2P3) grid sites.  Through the  JENNIFER2 project (funded by the Horizon 2020 programme\footnote{ The JENNIFER2 project is funded under the Horizon2020 program of the European Union as a Marie Sklodowska Curie Action of the RISE program:  MSCA-RISE-2018 call, project JENNIFER2, GA 822070}) and collaboration between the T2K, Hyper-K and Belle II~\cite{bib:belleII} experiments, Hyper-K connected cloud resources hosted by INFN and IN2P3 to the GridPP DIRAC instance using the virtual machine manager VCYCLE~\cite{bib:vcycle} to generate DIRAC pilot based VMs.  Following from this work, a cloud demonstrator was deployed that integrates token based authentication in the Openstack module of VCYCLE, allowing European Grid Infrastructure (EGI) cloud resources to be accessed by DIRAC~\cite{bib:jen2}.  This demonstrates versatile nature of DIRAC, facilitating the integration of cloud resources into the central pool for Hyper-K resources.

\subsubsection{Software Distribution}
\label{sec:softwareDist}
  All Hyper-K production is done with version controlled containers, which ensures consistent and reproducible results.  Hyper-K uses the CERN Virtual Machine File System (CVMFS)~\cite{bib:cvmfs} configured for EGI and hosted by RAL (UK), where these containers are distributed as singularity sandboxes.

%   All Hyper-K production is done with version controlled containers, which ensures consistent and reproducible results.  Thee containers are converted to singularity sanboxes and distributed onto the CERN Virtual Machine File System (CVMFS)~\cite{bib:cvmfs} that is configured for  European Grid Infrastructure (EGI) and hosted by RAL (UK) Singularity container sandboxes, that contain the Hyper-K software stack, are deployed onto CVMFS for the use of production jobs.

\subsubsection{Hyper-K Computing tools}
\label{sec:hkcomp}

The Hyper-K computing tools are written in python and designed to be platform-independent, such that a coherent set of tools and job definitions can be used across multiple resources, with a different wrapper written for each type of backend.  For submission to DIRAC-linked sites, the tools utilize the python-based DIRAC API.  Similarly, for all DFC data management, the Hyper-K tools integrate the DIRAC API to make use of the DMS and RMS tools for both single and bulk file operations and metadata management.  The DIRAC software may be installed locally, accessed on CVMFS or containerised with the Hyper-K tools, such that the minimum requirement from a site that wishes to use the full extent of the Hyper-K computing package is that it has open the necessary ports to access the GridPP DIRAC server\footnote{Though note that access to the DIRAC server is not necessary to use all aspects of the Hyper-K computing tools.}.

\subsection{Computing infrastructure and Tiers}
\label{sec:hkCompTiers}

\subsubsection{Grid/DIRAC Tiers}

This section outlines the Hyper-K tiers that are connected to the hyperk.org VO through the GridPP instance of DIRAC.  While this is primarily grid resources, it may also include cloud and batch systems, as mentioned in Section \ref{sec:dirac}.

\label{sec:GridDiracTiers}
\textbf{Tier-0 (T0)} sites are where raw data from the detectors is initially archived.  This is KEK Central Computer System (KEKCC) for the near detectors, and a dedicated computing system at the Kamioka Observatory for the far detector; both are based in Japan, close to the associated detector sites.  Raw data processing of each detector should be performed at the corresponding T0 site, although DIRAC-linked resources may be used as an over-spill if required, e.g. if a large amount of data needs to be reprocessed at short notice.  It is planned that 3,000 cores will be dedicated to this purpose at the Kamioka Observatory.  This is around double what is estimated for real-time processing of the HKFD, to allow for any reprocessing needs. For the near detectors, the CPU required for data processing is a small fraction of their MC production CPU needs.  However, to ensure data processing and reprocessing can be performed in a timely manner and at short notice, the number of cores allocated is based on being able to reprocess a years worth of beam data in around 2 months.  This will initially be set at around 1,000 cores, with the requirement increasing as more data is collected.  When these machines are not in use for data processing, these resources can be allocated for other purposes such as calibration or MC production.  

\textbf{Tier-1 (T1)} sites will hold copies of the raw and processed data, as well as MC production batches.  They will also generate and process most of the MC simulations.  The countries that are Tier-1 during the construction phase, and intend to maintain this status throughout operation, are France, Italy and the UK.  

\textbf{Tier-2 (T2)} sites provide disk storage that is mostly utilised in a temporary manner based on demand.  T2 sites can support the MC production campaigns with CPU, or may be used for specific tasks, especially in instances where the disk is utilized such that jobs can access the files locally.  Some or all of these countries that are T1 will also provide resources at T2 sites.  Other countries that are looking to secure grid resources for Hyper-K are Canada, Japan, Poland, Sweden and Switzerland.

\subsubsection{Standalone Tiers}
\label{sec:standaloneTiers}
The Hyper-K computing model is not confined to DIRAC-linked resources; some countries will contribute `standalone' computing resources such as local HTC and HPC clusters, cloud resources, and client-server storage solutions\footnote{Local clusters have the potential to be connected to DIRAC via SSH, but most will likely be treated as standalone.  Cloud resources may also be added to DIRAC; whether Hyper-K connects them to DIRAC or treats them as a standalone will depend on both the purpose of these resources and any technical considerations for a given site.}

\textbf{Standalone Compute-1 (SC1)} sites provides CPU/GPU at similar level of service as as T1; an example is the Kamioka Observatory computing system (which will provide both SC1-CPU and T0-Storage services), or the resources allocated to Hyper-K through The Digital Research Alliance of Canada.  

\textbf{Standalone Compute-2 (SC2)} resources are defined as a site that commits to providing the resources and person power to manage a specific computing service task; these are typically  institute clusters, and the Hyper-K computing group does not interact directly with these resources, only with the collaborator assigned to the task.  

\textbf{Standalone Storage-3 (SS3)} storage is any non-DIRAC/non-grid storage that is accessible to all Hyper-K collaborators.  It may be used for analysis files or general-purpose file sharing.  The KCL (UK) hosted Nextcloud service is an example of this.

%For tables use syntax in table~\ref{tab-1}.
%\begin{table}
%\centering
%\label{tab-1}       % Give a unique label
%% For LaTeX tables you can use
%\begin{tabular}{lll}
%%\hline
%first & second & third  \\\hline
%number & number & number \\\hline
%\end{tabular}
% Or use
%\vspace*{5cm}  % with the correct table height
%\end{table}
%

%

%%%%%%%%%%%%%%%%%%%%%%%%%%%%%
%%%%%%%%%%%%%%%%%%%%%%%%%%%%%
\section{Summary}
\label{sec:summary}
%%%%%%%%%%%%%%%%%%%%%%%%%%%%%

The Hyper-Kamiokande experiment has a rich and diverse physics programme, enabled by its suite of specialised, yet multi-purpose, detectors.  This gives rise to substantial computational and storage demands, met by significant infrastructure and pledges contributed by France, Italy, Japan, and the UK.  To coherently manage these contributions, and to incorporate a heterogeneous pool of resources from additional sites and platforms going forward, a centralised, efficient, and well-coordinated computing model is essential.  This is supported and realised by a combination of community software and services, notably DIRAC and the services provided by GridPP, and dedicated Hyper-K computing tools.  As a result, the Hyper-K computing group can enable measurements based on the latest data to be published in an efficient and timely manner, while ensuring the long-term preservation of data and metadata.

%%%%%%%%%%%%%%%%%%%%%%%%%%%%%
%%%%%%%%%%%%%%%%%%%%%%%%%%%%%
\section{Acknowledgements}
%%%%%%%%%%%%%%%%%%%%%%%%%%%%%
%%%%%%%%%%%%%%%%%%%%%%%%%%%%%

The author would like to thank the computing services and support provided by the following.  The Digital Research Alliance of Canada, GridPP, INFN-CNAF, the Centre de Calcul de l’IN2P3,  KEKCC, and Kamioka Observatory (ICRR, The University of Tokyo) for computing resources and services.  The European Union’s Horizon 2020 Research and Innovation Programme for funding the JENNIFER2 project that supported part of this work.  The Belle II collaboration, in particular S. Pardi, for collaboration on the cloud demonstrator.  The GridPP Collaboration and Imperial College London for the GridPP DIRAC services, in particular the support from D. Bauer and S. Fayer.  The CVMFS taskforce at UKRI RAL, funded by EGI.

%%%%%%%%%%%%%%%%%%%%%%%%%%%%%
%%%%%%%%%%%%%%%%%%%%%%%%%%%%%

\end{document}